\documentclass[prd,letterpaper,tightenlines,
showpacs,preprintnumbers,nofootinbib,floatfix,psfig]{revtex4}

\usepackage{amsmath,amssymb,amsfonts,graphicx,pifont,dcolumn}
\usepackage{epstopdf}
\usepackage{color}
\usepackage{cancel}
\usepackage{ulem}
\RequirePackage{slashed}

\begin{document}

\title{AdS gravity and  glueball spectrum}

\author{Vicente Vento} 
\address{Departament de Fisica Te\`orica, Universitat de Val\`encia
and Institut de Fisica Corpuscular, Consejo Superior de Investigaciones
Cient\'{\i}ficas, 46100 Burjassot (Val\`encia), Spain}

\date{\today}

\begin{abstract}
The glueball spectrum has attracted much attention since the formulation of Quantum Chromodynamics. Different approaches give very different results for their masses.
We revisit the problem from the perspective of the AdS/CFT correspondence. 
\end{abstract}


\maketitle

\section{Introduction}

Glueballs have been a matter of study and experimental search 
since the formulation of the theory of the strong interactions
Quantum Chromodynamics
(QCD)\cite{Fritzsch:1973pi,Fritzsch:1975wn}. QCD sum
rules \cite{Shifman:1978bx,Novikov:1979va}, QCD based models
\cite{Mathieu:2008me} and Lattice QCD computations, both with sea
quarks \cite{Gregory:2012hu} and in the pure glue theory
\cite{Morningstar:1999rf,Chen:2005mg,Meyer:2004jc} have been used to
determine their spectra and properties. However, glueballs have not been an easy 
subject to study due to the lack of phenomenological support and 
therefore much debate has been  associated with their properties \cite{Mathieu:2008me}. 
In particular, the lightestst scalar glueball has been a matter of discussion because depending on the approaches its calculated mass  ranges  from $m \sim 700$ MeV \cite{Dominguez:1986td,Bordes:1989kc,Mennessier:1985mz,Vento:2004xx} to that of lattice calculations $m > 1500$ MeV \cite{Gregory:2012hu,Morningstar:1999rf,Chen:2005mg,Meyer:2004jc}. Our idea here is to discuss what the AdS/CFT correspondence has to add to this controversy. In this letter we center our analysis in  the spectrum in Gluodynamics obviating the  difficult problem of mixing.

\begin{table}[htb]
\centering
\begin{tabular}{|c| c  c c c c c c  c|}
\hline 
&&  &  Quenched Lattice  & & &   &  Unquenched Lattice   & \\ \hline   
 & MP &  & CA& & MT &  &   GI    &    \\ \hline 
$0^{++}$ & $1730\pm 130$ &  &$1710 \pm 130$  &  & $1475 \pm 95$ &  &$1795 \pm 60 $&\\ \hline
$2^{++}$ & $2400 \pm 145$ &   & $2390 \pm150$ &  & $2150 \pm130$ &  & $2620 \pm 50$ & \\ \hline
$0^{++}$  & $2670 \pm 310$ &  &  &  & $2755\pm 150$ &  & $3760 \pm 240$  & \\ \hline
$0^{-+}$  & $2590 \pm 170$ &  & $2560 \pm 155$ &  & $2250\pm 160$ &  & $2887 \pm 180$ &\\ \hline
$1^{+-}$ & $2940 \pm 170$  & & $2980 \pm 170 $   &  & $2670 \pm 185$ &  &$3730 \pm 233 $ & \\ \hline
$1^{--}$ &$3850 \pm 240$ &   &  $3830 \pm 230 $ &  & $3240 \pm 480$  &  &$4658 \pm 291$& \\ \hline
\end{tabular}
\caption{Lattice glueball spectrum obtained from references MP \cite{Morningstar:1999rf}, CA \cite{Chen:2005mg}, MT \cite{Meyer:2004jc} and GI \cite{Gregory:2012hu}.}
\label{lattice}
\end{table}

\section{The glueball spectrum}

We recall the glueball spectrum of lattice QCD  in Table \ref{lattice}. 
A phenomenological analysis of the lattice spectrum at the light of the $f_0$ spectrum leads to the conclusion that a scalar glueball should exist in quenched QCD  in the mass range $1650-1750$ MeV \cite{Vento:2015yja}. Let us isolate the quenched lattice glueball spectrum as shown in Table \ref{lowglue}. We note several features.  The lightest glueball is a scalar $0^{++}$ with a mass on average $\sim 1600$ MeV with errors at the level of $100$ MeV \cite{Morningstar:1999rf,Chen:2005mg,Meyer:2004jc}, precisely $1638 \pm 119$ MeV.  The two closest excitations are  a $2^{++}$ and $0^{++}$   with masses $2331 \pm 142$ and $2712 \pm 244$ MeV respectively. Given their large errors we could in principle consider the latter two to first approximation degenerate. In summary we will demand from our spectrum to have a low mass $0^{++}$ scalar and some  $2^{++}$, $0^{++}$ degenerate pair. One should also notice that lattice calculations misses particles. For example if we look at Table \ref{lattice} we see in the second column that the excited $0^{++}$ is missing and in the fourth column the $0^{-+}$ and the $1^{+-}$ have lower masses than the  $0^{++}$, an indication that some states might be missing. With these caveats we proceed to our analysis of the AdS/CFT spectrum aiming at reproducing these characteristics.

\section{The AdS/CFT glueball spectrum}

The AdS/CFT correspondence provides new techniques to deal with non abelian gauge theories.  The Maldacena duality conjecture \cite{Maldacena:1997re} and subsequent developments \cite{Witten:1998qj,Witten:1998zw,Gubser:1998bc}  lead to a geometrical picture for QCD which is based on an $AdS_7$ soliton whose metric is \cite{Csaki:1998qr,Constable:1999gb,Brower:2000rp}

\begin{equation}
ds^2 =(r^2-\frac{1}{r^4}) d\tau^2 +r^2 \eta_{\mu \nu} dx^\mu dx^\nu +(r^2 - \frac{1}{r^4})^{-1} ) dr^2
\label{metric}
\end{equation}
where $\eta_{\mu\nu}$ is the Minkowski metric in five dimensions. 

 The strong coupling glueball calculation consists in finding the normal modes for the supergraviton multiplet. The supergravity modes represent excitations of QCD operators which posses a mass spectrum. One has to find all quadratic fluctuations in the background that survive  for QCD in the scaling limit. The result in appropriate units is given in Table \ref{adsmodes}. Note that in the calculation there are two sources of $0^{++}$ states one is associated with the dilaton field and the other with the scalar component of the AdS graviton. The latter is the lightest one. Moreover, the former, the dilaton, is degenerate with the tensor component of the AdS graviton \cite{Csaki:1998qr,Brower:2000rp}. In order to move from the AdS modes to the glueball spectrum we need a scale.  To fix the scale we use the assumed approximate degeneracy between the $2^{++}$ and the $0^{++}$ described above and which arises naturally in the AdS/CFT result as seen in Table \ref{adsmodes}. We thus assume that in the spectrum the $2^{++}$ and the first excited $0^{++}$ are degenerate with a mass between $2300-2700$ MeV and we choose these degenerate pair fo fix the scale. We study two schemes : the first scheme assumes that a mass of $2300-2700$ MeV corresponds to the first degenerate pair of the AdS/CFT spectrum; the second scheme assumes that this mass corresponds to the second degenerate pair of the AdS/CFT spectrum. The result of that study is shown in Table \ref{adsspectrum}.
 \begin{table}[htb]
\begin{center}
\begin{tabular}{|c| c c c|}
\hline
$J^{PC}$  &   MP & CA & MT \\  \hline
$0^{++}$ &  1$730 \pm 130 $& $1710 \pm 130 $& $1475 \pm  95$ \\
$2^{++}$ & $ 2400 \pm 145$ & $ 2390 \pm 150$ & $2150\pm 130$ \\
$0^{++}$ & $2670 \pm 310$ & &  $2755 \pm150 $ \\ 
\hline
  \end{tabular}
  \label{lowglue}
\end{center}
\caption{Glueball masses with $J^{PC}$ assignments.
The columns MP~\cite{Morningstar:1999rf}, KY~\cite{Chen:2005mg}  and MT~\cite{Meyer:2004jc}}
\label{gmass}
\end{table}

The first fit is called AdS1 in Table \ref{adsspectrum} and its lightest resonance is a scalar glueball  whose mass is between $1300-1500$ MeV, which is low compared with the lattice average.  Other resonances are: the $0^{-+}$ in  good agreement  with the lattice; the $1^{+-}$ and $1^{--}$ which are too massive; additional $0^{++},2^{++},0^{++},0^{-+}$ states in the intermediate range which do not appear in the lattice spectrum. A caveat, the second $0^{++}$ state of the unquenched calculation \cite{Gregory:2012hu} has a mass which is close  to that of the additional  $0^{++}$ state, which may hint  missing resonances in the quenched calculations. Finally, if we consider that the AdS calculation is a large $N$ approximation and we look into lattice calculations dealing with large $N$ as shown in Table \ref{largeN} \cite{Lucini:2001ej,Lucini:2004my}, we notice that to compare with the $SU(3)$ lattice calculations we must increase the scalar masses by $10\%$, while the tensors do not change with $N$. This would fix the light scalars and tensors but not the high lying resonances, and moreover the AdS spectrum remains too crowded unless our caveat is confirmed. In any case we conclude from this analysis that given the simplicity of the AdS model this fit supports the QCD lattice spectrum.

\begin{table}[htb]
\centering
\begin{tabular}{|l| c|  c| c| c| c|}
\hline 
$J^{PC}$ & $ 0^{++} $ & $ 2^{++}/1^{++}/0^{++} $ & $1^{- +}/0^{- +}$ & $1^{+ -}/0^{+ -} $ & $ 1^{- -}/1^{- -} $ \\
\hline
n=1 &$7.308$ &$ 22.097$ & $31.985$ &$53.376$&$ 83.046$ \\
\hline
n=2 &$ 46.986$ & $55.584$ & $72.489$ & $109.446$ & $143.582$ \\
\hline
\end{tabular}
\caption{The mode spectrum of the supergraviton , $m_n^2$ for QCD glueballs from ref. \cite{Brower:2000rp}.}
\label{adsmodes}
\end{table}

The second fit called AdS2 is based on fixing the masses to the degeneracy of the second pair. It provides new ingredients with respect to the lattice results: a  low mass scalar ($m \sim 900$ MeV) not seen in lattice calculations;  the seen scalar at $m \sim1600$ MeV, but in this fit degenerate with an unseen tensor; many states close to the fitted degenerate pair which pile up closer than in lattice calculations; the masses of the higher lying resonances are in reasonable agremment with lattice results. This fit supports the approaches which predict a low mass scalar glueball. Given the crudeness of the AdS model we  argue that the most crucial detectable signature of the fit is the doubling of the $ \sim 1600$ glueball with the tensor. If this doubling is found in lattice calculations then necessarily the low mass glueball should be looked for.  The  corrections for large $N$  do not alter these statements. 

The naive AdS model used might not be precise in getting the mass levels and orderings but it seems consistent  with the labelling of quantum numbers. Therefore the existence of a closely lying $2^{++}$ tensor to the $0^{++}(1600)$ glueball would be a signature of AdS2.

\begin{table}[htb]
\centering
\begin{tabular}{|l| l  c  l|}
\hline 
 & AdS1 &   & AdS2  \\ \hline 
$0^{++} $ & 1323 - 1553 && $\;\,834 -\;979$ \\ \hline
$2^{++}$ & 2300 - 2700*&&1450 - 1702 \\ \hline
$0^{++}$ &2300 - 2700*&& 1450 - 1702 \\ \hline
$0^{-+}$& 2767 - 3248 & &1744 - 2048\\ \hline
$0^{++}$ & 3353 - 3937&&2114 - 2482 \\ \hline
$1^{+-}$ &3575 - 4196&&2253 - 2645   \\ \hline
$2^{++}$ &3648 - 4282& &2300 - 2700*\\ \hline
$0^{++}$ &3648 - 4242& &2300 - 2700*\\ \hline
$0^{-+}$&4165 - 4890 & &2626 - 3083 \\ \hline
$1^{--}$& 4449 - 5234& &2811 - 3300 \\ \hline
 
\end{tabular}
\caption{Gueball spectrum for two  parametrizations of the AdS modes of  ref. \cite{Brower:2000rp} obtained by fixing the scale as indicated in the text. The * signals the data used to  fix the scale.}
\label{adsspectrum}
\end{table}

\section{Concluding remarks}

We have revisited the AdS/CFT glueball spectrum at the light of new phenomenological analyses of the scalar glueballs and QCD lattice calculations with the aim of clarifying the controversy regarding the mass of the lightests glueball. In order to do so we have taken the liberty of fixing the scale to analyze two possible scenarios which reproduce the two conflicting views. Under the working assumption that some states might be missing in lattice calculations we have found that according to AdS/CFT a light scalar glueball carries unavoidably to an almost degenerate $2^{++}$ resonance to the  $0^{++}(1600)$  glueball. Despite the simplicity of the model used, which might affect the precise values of the masses, the labelling of the quantum number of the states and their degeneracies seem to be consistent with the QCD spectrum and this gives us a leeway to distinguish between the two scenarios.

\begin{table}[htb]
\centering
\begin{tabular}{|c| c  c c c  c|}
\hline 
  
  & $0^{++}$ &  & $0^{++*}$ & & $ 2^{++}$\\ \hline 
  & &  &  Continuum   &  &  \\ \hline 
 m(SU(3))/m(SU($\infty$)) & $1.07\pm 0.04$ &   &$0.94  \pm 0.04$  &  & $1.00 \pm 0.03$ \\ \hline
  & & & Smallest lattice   & &  \\ \hline   
 & $1.17 \pm 0.05 $&  & $1.00 \pm 0.04$ & & $0.99 \pm 0.04$\\ \hline
  & & & Average&   &  \\ \hline   
   & $1.12 \pm 0.03 $&  & $0.97 \pm 0.03$ & & $1.00 \pm 0.03$\\ \hline
\end{tabular}
\caption{Ratios of glueball masses for N=3 and very large N as shown in ref. \cite{Lucini:2001ej,Lucini:2004my}.}
\label{largeN}
\end{table}

\section*{Acknowledgments}

This work was supported in part by Mineco under contract  FPA2013-47443-C2-1-P, Mineco and UE Feder under contract FPA2016-77177-C2-1-P, GVA- PROMETEOII/2014/066 and SEV-2014-0398.

\end{document}